\documentclass[12pt,fleqn]{article}
\usepackage{xiiiemc}
\usepackage{natbib}
\usepackage{fancyhdr}
\usepackage{fancyvrb}
\usepackage{color}
\usepackage{wallpaper} 
\usepackage{titlesec}   
\usepackage{float} 	
\usepackage[usenames,dvipsnames]{xcolor}
\usepackage{hyperref} 	
\hypersetup{
	pdfcreator={LaTeX with abnTeX2},
	pdfkeywords={abnt}{latex}{abntex}{USPSC}{trabalho acadêmico}, 
	colorlinks=true,       		
	linkcolor=blue,          	
	citecolor=blue,        		
	filecolor=magenta,      		
	urlcolor=blue,
	allbordercolors=black,
	bookmarksdepth=4
}
\usepackage{tocloft}
\titleformat{\section}
  {\normalfont\bfseries}{\thesection.}{0.5em}{}
 
\renewcommand\thesection{\arabic{section}}

\let\OLDthebibliography\thebibliography
\renewcommand\thebibliography[1]{\OLDthebibliography{#1} \setlength{\parskip}{0pt}\setlength{\itemsep}{0pt plus 0.3ex}}

\usepackage{grffile}
\usepackage{afterpage}

\title{A STATISTICAL DISTANCE DERIVED FROM THE KOLMOGOROV-SMIRNOV TEST: SPECIFICATION, REFERENCE MEASURES (BENCHMARKS) AND EXAMPLE USES}

\author
    {\rm \begin{tabular}{l} 
	    \textbf{Renato Fabbri}$^{1}$ - {\textnormal renato.fabbri@gmail.com}\\%
    \textbf{Fernando Gularte De León}$^{2}$ - {\textnormal fernandogularte@gmail.com}\\
	    {\fontsize{11}{0}\selectfont $^{1}$University of São Paulo, Institute of Mathematical and Computer Sciences - São Carlos, SP, Brazil}\vspace*{-0.05cm} \\
	    {\fontsize{11}{0}\selectfont $^{2}$Parque Tecnológico Industrial del Cerro - Montevideo, MO, Uruguay}\vspace*{-0.05cm}\\
  \end{tabular}}

\fancypagestyle{firspagetstyle}
{
	\lhead{}
	\fancyhead[C]{%
		\includegraphics[width=0.9\linewidth]{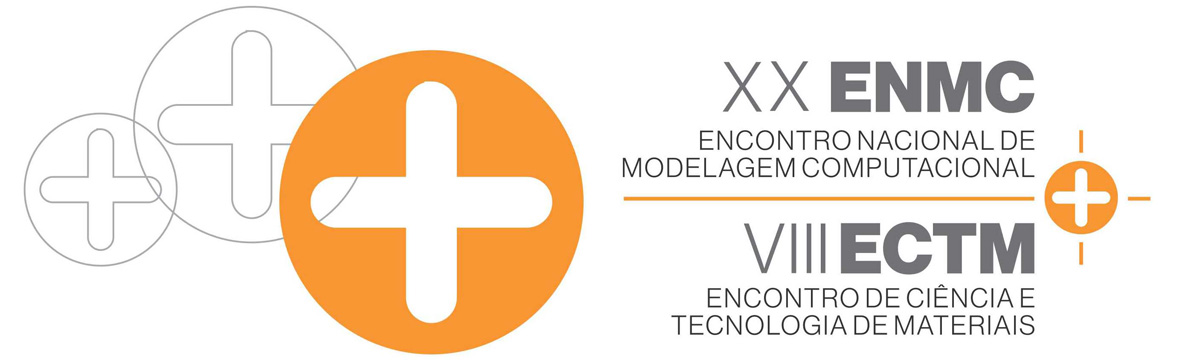}\\%
		{\scriptsize \fontfamily{phv}\fontseries{b}\selectfont \color[rgb]{0.45,0.45,0.45}
		16 a 19 de Outubro de 2017\\
		Instituto Politécnico - Universidade do Estado de Rio de Janeiro\\
		Nova Friburgo - RJ\\
	    }
	}
	\renewcommand{\headrulewidth}{0.0pt}
	\fancyfoot[C]{\footnotesize \parbox{15cm} {\centering  \fontsize{7.5}{0}\selectfont \it Anais do XX ENMC – Encontro Nacional de Modelagem Computacional e VIII ECTM – Encontro de Ciências e Tecnologia de Materiais,  Nova Friburgo, RJ – 16 a 19 Outubro 2017}} 
	\rhead{}
}

\begin{document}
\maketitle

\thispagestyle{firspagetstyle}

\fancyhead[L]{\footnotesize{\fontsize{7.5}{0}\selectfont \it XX ENMC e VIII ECTM\\
	16 a 19 de Outubro de 2017\\
	Instituto Politécnico Universidade do Estado do Rio de Janeiro – Nova Friburgo - RJ\\}}
\renewcommand{\headrulewidth}{0.0pt}
\fancyfoot[C]{\footnotesize \parbox{15cm} {\centering  \fontsize{7.5}{0}\selectfont \it Anais do XX ENMC – Encontro Nacional de Modelagem Computacional e VIII ECTM – Encontro de Ciências e Tecnologia de Materiais,  Nova Friburgo, RJ – 16 a 19 Outubro 2017}} 
\rhead{}

\begin{abstract}
Statistical distances quantifies the difference between two statistical constructs.
In this article, we describe reference values for a distance between samples
derived from the Kolmogorov-Smirnov statistic $D_{F,F'}$.
Each measure of the $D_{F,F'}$ is a measure of difference between two samples.
This distance is normalized by the number of observations in each sample
to yield the $c'=D_{F,F'}\sqrt{\frac{n n'}{n+n'}}$ statistic,
for which high levels favor the rejection of the
null hypothesis (that the samples are drawn from the same distribution).
One great feature of $c'$ is that it inherits the robustness of
$D_{F,F'}$ and is thus suitable for use in settings where
the underlying distributions are not known.
Benchmarks are obtained by comparing samples derived from standard distributions.
The supplied example applications of the $c'$ statistic for the distinction
of samples in real data enables further
insights about the robustness and power of such statistical distance.
\end{abstract}


\pagestyle{fancy}

\section{INTRODUCTION}\label{sec:intro}
To quantify the difference between samples that are regarded as statistical events,
one can rely in statistical distances.
Such distances are often not metrics, cases in which they do not satify one or more of
the properties of a metric $m$ on samples $x_i \in X$:
\begin{align}
	m(x_i,x_j) &  \geq 0 \\
	m(x_i,x_j) &  = 0 \Leftrightarrow x_i = x_j \\
	m(x_i,x_j) &  = m(x_j,x_i)\\
	m(x_i,x_j) &  \leq m(x_i,x_z) + m(x_z,x_j)
\end{align}

Pseudometrics violate property (1) and/or (2),
quasimetrics violate property (3),
semimetrics violate property (4).
A divergence only satisfies properties (1) and (2).
These are ``generalized metrics''~\citep{wikiStatDist}.

In this article, a statistical distance $c'$ derived from the
Kolmogorov-Smirnov test is described.
The $c'$ statistic can be both a true or a generalized metric,
depending on the implementation details, as explained in Section~\ref{sec:desc}.
To enable the use of the $c'$ metric,
benchmarks are provided
by using standard distributions in various settings and sample sizes.
Example applications of the metric to quantify the difference among
real signals further validate the approach.

Section~\ref{sec:met} describes the metric
and the methods used to characterize it.
Section~\ref{sec:res} is dedicated to
summarizing the results and essential discussions.
Final remarks, including potential future works,
are stated in Section~\ref{sec:conc}.

\section{METHODS}\label{sec:met}
This section describes the $c'$ statistical distance,
the strategy of benchmarking and the validation of $c'$ by means
of application to real samples.

\subsection{Description of the $c'$ statistic}\label{sec:desc}
\begin{figure}[!htbp] 
\vspace{-2pt}
\begin{center}
	\includegraphics[width=0.44\textwidth]{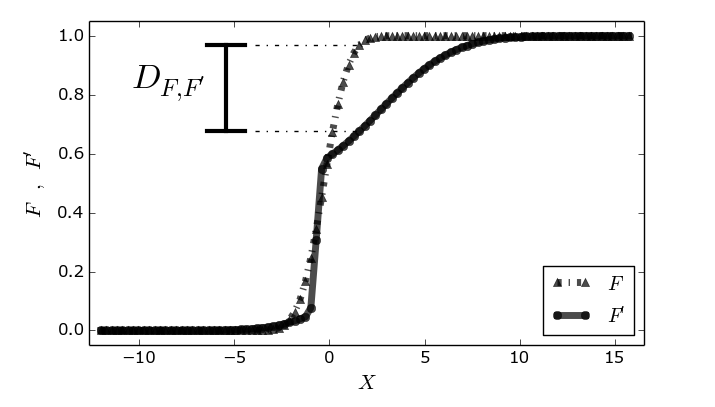}
	\caption{The Kolmogorov-Smirnov statistic $D_{F,F'}$: the maximum difference between
		two cumulative distribution functions.}
	\label{fig:dnn}
\end{center}
\end{figure}

Be $F$ and $F'$ two empirical cumulative distributions,
where $n$ and $n'$ are the number of observations in each sample.
The two-sample Kolmogorov-Smirnov test rejects the null hypothesis,
that the histograms are the outcome of the same underlying distribution,
if:
\begin{equation}\label{eq:ks}
D_{F,F'} > c(\alpha)\sqrt{\frac{n+n'}{nn'}}
\end{equation}

\noindent where $D_{F,F'}=sup_x[F-F']$ as in Figure~\ref{fig:dnn}
and $c(\alpha)$ is related to the level of significance $\alpha$ by:

\begin{table}[h!]
\centering
\begin{tabular}{|l||c|c|c|c|c|c|}\hline
$\alpha$    & 0.1  & 0.05 & 0.025 & 0.01 & 0.005 & 0.001 \\\hline
$c(\alpha)$ & 1.22 & 1.36 & 1.48  & 1.63 & 1.73  & 1.95  \\\hline
\end{tabular}
\end{table}

If distributions are drawn from empirical data, $D_{F,F'}$ is given as are $n$ and $n'$.
All terms in equation~\ref{eq:ks} are positive and $c(\alpha)$ can be isolated:

\begin{equation}\label{eq:ks2}
	c(\alpha) < D_{F,F'}\sqrt{\frac{nn'}{n+n'}} = c'
\end{equation}

The higher $c'$ is, the lower $\alpha$ can be and still entail the rejection of the null hypothesis.

In summary,
high values of $c'$ favor rejecting the null hypothesis.
For example, if the significance level is $\alpha=0.01$,
then $c'$ greater than $1.7$
implies the rejection of the null hypothesis,
i.e. implies the assumption that $F$ and $F'$
are outcomes of different distributions.

Of core importance in this study is to regard $c'$
as a measure of distance between both distributions.
If fact, it is a statistical distance.
Following the concepts defined in Section~\ref{sec:intro},
it can be both a true or a generalized metric, depending on the implementation.
It obviously satisfies the Equations (1) and (3).
It satisfies Equation (4) for less obvious reasons.
To grasp how $c'$ satisfies Equation (4),
let $x_i$, $x_j$ and $x_z$ be samples of the same size,
so we only need to compare the $D_{F,F'}$.
Let $F_i$ be the cumulative distribution of the sample $x_i$.
Supose that in the value $\xi$ of X where $F_i$ and $F_j$ are maximally different (i.e. where they yield $D_{F_i,F_j}$),
they are also maximaly different against $F_z$.
If the value of $F_z(\xi)$ is between $F_x(\xi)$ and $F_j(\xi)$:
$D_{F_i,F_z}+D_{F_j,F_z} = D_{F_i,F_j}$,
otherwise:
$D_{F_i,F_z}+D_{F_j,F_z} > D_{F_i,F_j}$.
If the KS statistic $D_{F,F'}$ are not yield at the same value,
it is because they are larger than in the previous cases, thus:
$D_{F_i,F_z}+D_{F_j,F_z} > D_{F_i,F_j}$.
And this completes the argument for:
$D_{F_i,F_z}+D_{F_j,F_z} \geq D_{F_i,F_j}$.
The $c'$ statistic might satisfy or violate Equation (2),
depending on how it is achieved.
If the obtainance of $c'$ depends on making histograms,
than a slightly different observation of a sample might fall under the same bin.
In this case, $c'=m(x_i,x_j) = 0$ and $x_i \neq x_j$, which violates (2)\footnote{If
we instead regard $x_i$ as a histogram, not a sample,
then it satifies (2), but we here assume that $c'$
is in fact a measure related to samples.}.
The cumulative distributions might be derived, however,
not by making a histogram, but simply by ordering the samples~\citep{stack}.
In this case, $c'$ satifies (2).
One exception: if $x_j$ has twice each of the observations in $x_i$,
then it violates (2)
because the distributions entailed by the samples are the same,
but the samples are not the same, and the distance is still zero.

In summary, if $c'$ can be classified both as a metric and a pseudometric,
depending on how it is obtained and theoretical nuances.

\subsection{Benchmarks obtainance}
We considered two cases: when the null hypothesis (that the samples were drawn from the same underlying distributions)
is true and when it is false.
For the case where the null hypothesis is true, 
we compared similar distributions in various settings
many times to assert that we would not assume
that the null hypothesis was false more than
$\alpha . N_c$ where $\alpha$ is the significance level
and $N_c$ is the number of comparisons.
That is, to assert that the Kolmogorov-Smirnov test results
are in accordance with the theory.
In the case where the null hypothesis was false, 
we were interested in measures of $c'$ given that
the null hypothesis is never rejected for a small enough
$\alpha$.
The various measures performed for $c'$ are
described in the results.

One important aspect of the way by which we made the
benchmarks available is that the rendering of the tables
is automated by configurable scripts,
allowing one to obtain tables with other measures
and other comparisons.

\section{RESULTS AND DISCUSSION}\label{sec:res}
This section briefly describes each of the
results, which are: benchmark tables, example uses of $c'$ in
real samples, an exposition of all the data obtained,
and configurable scripts for the generation of all reference tables.

\subsection{When the null hypothesis is true}\label{sec:true}
The theory of the Kolmogorov-Smirnov test
states that one can choose a significance level $\alpha$,
which is the maximum probability that one will reject the
null hypothesis when it is true.
Accordingly,
we rendered tables for each of the distributions:
normal, uniform, 1-parameter Weibull, power function.
Three to five different settings of each of the distributions
were used, both samples had a size of 1000 observations,
and $N_c=100$ comparisons were performed.

Table~\ref{tab:true} is one of such tables.
To understand the columns, notice that
if the null hypothesis is true, the number
of rejections of the null hypothesis ($c'>c(\alpha)$)
in $N_c$ comparisons should not exceed $\alpha . N_c$.
To verify this, let $C=\{c'_i\}$ be a set of $c'$ measures,
and $C(\alpha)=\{c' : c'>c(\alpha)\}$.
Be $|C(\alpha)|$ the cardinality of $C(\alpha)$,
i.e. the number of comparisons in which the two-sample Kolmogorov-Smirnov
test rejects the null hypothesis for a given $\alpha$.

The overall result is that, in fact, the false rejections of the null hypothesis
does not exceed $\alpha . N_c$.
The only exception in our simulations is the power-law (or power function) distribution,
in which the number of rejections of the null hypothesis were usually bellow $\alpha . N_c$
but, in extreme cases, our simulations reached almost $2\alpha N_c$.

\begin{table}[H]
\caption{The theoretical maximum number $\alpha N_c$ of rejections
        of the null hypothesis for the critical value $\alpha$.
        The $c'_1$ values were calculated using simulations of 1-parameter Weibull distributions with $a=0.1$.
        The $c'_2$ values were calculated using simulations of 1-parameter Weibull distributions with $a=2$.
        The $c'_3$ values were calculated using simulations of 1-parameter Weibull distributions with $a=4$.
        The $c'_4$ values were calculated using simulations of 1-parameter Weibull distributions with $a=6$.
        Over all $N_c$ comparisons,
         $\mu(c'_1)=0.1107$ and $\sigma(c'_1)=0.0652$,
         $\mu(c'_2)=0.8079$ and $\sigma(c'_2)=0.2417$,
         $\mu(c'_3)=0.7775$ and $\sigma(c'_3)=0.2404$,
         $\mu(c'_4)=0.8209$ and $\sigma(c'_4)=0.2389$.
	 For more information on interpreting these measurements, see Section~\ref{sec:true}.
	}\label{tab:true}
\begin{center}
\begin{tabular}{l | c | c || c | c | c | c}
$\alpha N_c$ & $\alpha$ & $c(\alpha)$ & $|C_1(\alpha)|$ & $|C_2(\alpha)|$ & $|C_3(\alpha)|$ & $|C_4(\alpha)|$ \\\hline
10.0 & {\bf 0.100} & 1.22 & 0 & 9 & 5 & 9 \\
5.0 & {\bf 0.050} & 1.36 & 0 & 1 & 3 & 3 \\\hline
2.5 & {\bf 0.025} & 1.48 & 0 & 0 & 1 & 1 \\
1.0 & {\bf 0.010} & 1.63 & 0 & 0 & 1 & 0 \\\hline
0.5 & {\bf 0.005} & 1.73 & 0 & 0 & 1 & 0 \\
0.1 & {\bf 0.001} & 1.95 & 0 & 0 & 0 & 0 \\
\end{tabular}
\end{center}
\end{table}

\subsection{When the null hypothesis is false}\label{sec:false}
In this case we are interested in measures of $c'$.
The number of comparisons is still $N_c=100$.
The measures on $c'$ chosen to report the results are:
the mean $\mu(c')$, the standard deviation $\sigma(c')$,
the median $m(c')$,
the  fraction
$\overline{C(\alpha)}=\frac{|C(\alpha)|}{N_c}$
of rejection of the null hypothesis given the significance level $\alpha$,
$min(c')$ which states the three smallest values found in the simulations while
$max(c')$ states the three greatest values.
The null hypothesis is true in the boldface lines.
$D$ is the KS statistic when sample size is very large.

Two sets of tables were made to study the $c'$ statistic when
the null hypothesis is false:
\begin{itemize}
	\item Changing the distributions: in each table, the comparisons were made with
one of the distributions remaining unchanged
while the other changes in each row.
Table~\ref{tab:false1} is an example of such table.
	\item Changing the sample sizes:
		changing the number of elements in each sample
		changes the value of the $c'$ statistic. 
		Thus, $c'$ is given for two samples of varied sizes but
		with fixed underlying distributions.         
		Table~\ref{tab:false2} is an example of such table.
\end{itemize}

\begin{table*}[h!]
\caption{Measurements of $c'$ through simulations
        with power function distributions.
        One power distribution has the fixed exponent parameter $a-1=0.5$.
        The other power function distribution
	has a different value of $a$ in each row (i.e. in each set of $N_c$ comparisons).
	A description of the measures is in Section~\ref{sec:false}.
	}\label{tab:false1}
\tiny
\begin{center}
\begin{tabular}{ l | c | c | c | c | c | c | c | c | c | c | c }
$a$ & $\mu(c')$ & $\sigma(c')$ & min(c') & max(c') & $D$ & $\mu(D_{F,F'})$ & $\sigma(D_{F,F'})$ & $\overline{C(0.1)}$ & $\overline{C(0.05)}$ & $\overline{C(0.01)}$ &  $\overline{C(0.001)}$ \\\hline
0.7 & 6.282 & 0.402 & 4.919,5.299,5.501 & 6.909,6.999,7.021  & 0.274  & 0.281  & 0.018  & 1.000  & 1.000 & 1.000  & 1.000 \\
0.9 & 4.445 & 0.452 & 3.511,3.600,3.622 & 5.344,5.367,5.590  & 0.186  & 0.199  & 0.020  & 1.000  & 1.000 & 1.000  & 1.000 \\\hline
1.1 & 2.818 & 0.443 & 1.588,1.744,1.945 & 3.734,3.846,4.137  & 0.114  & 0.126  & 0.020  & 1.000  & 1.000 & 0.990  & 0.970 \\
1.3 & 1.536 & 0.407 & 0.783,0.827,0.894 & 2.415,2.437,2.549  & 0.053  & 0.069  & 0.018  & 0.750  & 0.650 & 0.350  & 0.170 \\\hline
{\bf 1.5} & {\bf 0.776} & {\bf 0.237} & {\bf 0.425,0.447,0.470} & {\bf 1.409,1.409,1.565} & {\bf 0.000} & {\bf 0.035} & {\bf 0.011} & {\bf 0.060} & {\bf 0.040} & {\bf 0.000} & {\bf 0.000} \\\hline
1.7 & 1.499 & 0.377 & 0.648,0.738,0.850 & 2.281,2.326,2.370  & 0.046  & 0.067  & 0.017  & 0.740  & 0.660 & 0.380 & 0.110 \\\
1.9 & 2.246 & 0.400 & 0.939,1.386,1.521 & 2.952,3.063,3.309  & 0.087  & 0.100  & 0.018  & 0.990  & 0.990 & 0.940 & 0.780 \\\hline
2.1 & 3.051 & 0.395 & 2.393,2.393,2.415 & 3.846,3.891,3.913  & 0.123  & 0.136  & 0.018  & 1.000  & 1.000 & 1.000 & 1.000 \\\
2.3 & 3.710 & 0.442 & 2.683,2.795,2.907 & 4.696,4.718,4.785  & 0.156  & 0.166  & 0.020  & 1.000  & 1.000 & 1.000 & 1.000 \\\hline
2.5 & 4.415 & 0.422 & 3.019,3.175,3.287 & 5.121,5.188,5.210  & 0.186  & 0.197  & 0.019  & 1.000  & 1.000 & 1.000 & 1.000 \\
\end{tabular}
\end{center}
\end{table*}
\begin{table*}[h!]
\caption{Measurements of $c$ through simulations
        with fixed normal distributions but different number of samples.
        One normal distribution has $\mu=0$ and $\sigma=1$.
        The other normal distribution has
        $\mu=0$ and $\sigma=1.2$.
        The KS statistic of these distributions converges
        to 0.0440 as sample sizes increases.
	A description of the measures is in Section~\ref{sec:false}.
	}\label{tab:false2}
\tiny
\begin{center}
\begin{tabular}{ l | c | c | c | c | c | c | c | c | c | c }
$n=n'$ & $\mu(c')$ & $\sigma(c')$ & $m(c')$ & $min(c')$ & $max(c')$ & $\mu(D_{F,F'})$ & $\sigma(D_{F,F'})$ & $\overline{C(0.1)}$ & $\overline{C(0.01)}$ & $\overline{C(0.001)}$ \\\hline
{\bf 100} & 0.911 & 0.251 & 0.849 & 0.424,0.495,0.495 & 1.414,1.626,1.697 & 0.129 & 0.035 & 0.130 & 0.010 & 0.000 \\
{\bf 1000} & 1.466 & 0.260 & 1.431 & 0.917,1.051,1.051 & 1.990,2.080,2.214 & 0.066 & 0.012 & 0.820 & 0.270 & 0.050 \\\hline
{\bf 10000} & 3.467 & 0.243 & 3.465 & 2.878,2.970,2.991 & 3.946,4.080,4.094 & 0.049 & 0.003 & 1.000 & 1.000 & 1.000 \\
{\bf 100000} & 10.129 & 0.253 & 10.125 & 9.595,9.595,9.631 & 10.713,10.735,10.896 & 0.045 & 0.001 & 1.000 & 1.000 & 1.000 \\
\end{tabular}
\end{center}
\end{table*}

\subsection{Example application to real samples}
To further validate the $c'$ statistic and enable deeper insights,
a number of applications to real samples were performed:
\begin{itemize}
	\item Texts: Hamlet (Shakespeare), the Bible (KJV), Moby Dick (Herman Melville) and Esaú e Jacó (Machado de Assis),
		where studied by regarding the stopwords and the words which were not stopwords.
		Each of these works were considered as a whole and divided in the first and second half.
                These texts were used to obtain samples that are:
		the mean of the token sizes, the standard variation of the token sizes,
		the token sizes.
		For the two first samples, the text was divided into 1000 parts in which the means
		and standard variations were obtained and regarded as the observations.
		The overall result is: smaller $c'$ for comparisons between parts of the same text
		although high $c'$ was incident even between parts of the same book (especially for the Bible,
		probably because of great differences between the New and Old Testaments).
	\item Audio: the audio segments for testing the sound system of an Ubuntu Linux distribution
		were considered both by their PCM samples and by their Daubechies 8 wavelet coefficients.
		The segments yielded higher $c'$ values as the audio held greater differences, e.g. yield by different words
		or noise.
	\item Music: each classical composition was regarded as a sample
		and the pitches were regarded as observations.
		The results reflect music history.
		For example, measures of $c'$ involving Palestrina
		increases along time with the exception of Beethoven
		who, indeed, used modalism.
		The values of $c'$ related to Bach also increases along time,
		and the outcome of the comparison against Palestrina 
		is only exceeded when Sch\"onberg is reached,
		which reflects the non-tonal discourse of both
		Palestrina and Sch\"onberg.
		Table~\ref{tab:mus} exposes these results.
	\item OS status: workload of the CPUs and memory allocation of the most consuming processes.
		Again, the type of samples are mandatory:
		they might all be identified by the values of $c'$
		found in comparison against other samples,
		with the exception of the RAM memory.
\end{itemize}

\begin{table}[h!]
	\caption{Values of $c'$ for pitches of classical compositions.}\label{tab:mus}
\small
\begin{center}
\begin{tabular}{ l | c || c | c || c | c || c | c || c }
 & {\bf Pale} & {\bf Bach1} & {\bf Bach2} & {\bf Moza1} & {\bf Moza2} & {\bf Beet1} & {\bf Beet2} & {\bf Sch\"on} \\\hline
{\bf Pale} & 0.00 & 1.88 & 1.89 & 2.60 & 4.12 & 4.43 & 5.49 & 2.62 \\
 & 0.00  & 0.14  & 0.13  & 0.15  & 0.17  & 0.21  & 0.21  & 0.28 \\\hline\hline
{\bf Bach1} & 1.88 & 0.00 & 1.00 & 1.27 & 1.50 & 2.09 & 2.51 & 1.54 \\
 & 0.14  & 0.00  & 0.09  & 0.10  & 0.10  & 0.15  & 0.16  & 0.18 \\\hline
{\bf Bach2} & 1.89 & 1.00 & 0.00 & 1.26 & 1.78 & 2.20 & 2.73 & 1.52 \\
 & 0.13  & 0.09  & 0.00  & 0.09  & 0.11  & 0.15  & 0.17  & 0.18 \\\hline\hline
{\bf Moza1} & 2.60 & 1.27 & 1.26 & 0.00 & 2.14 & 2.08 & 2.25 & 1.79 \\
 & 0.15  & 0.10  & 0.09  & 0.00  & 0.10  & 0.11  & 0.10  & 0.19 \\\hline
{\bf Moza2} & 4.12 & 1.50 & 1.78 & 2.14 & 0.00 & 2.99 & 5.52 & 2.02 \\
 & 0.17  & 0.10  & 0.11  & 0.10  & 0.00  & 0.10  & 0.09  & 0.20 \\\hline\hline
{\bf Beet1} & 4.43 & 2.09 & 2.20 & 2.08 & 2.99 & 0.00 & 2.34 & 2.31 \\
 & 0.21  & 0.15  & 0.15  & 0.11  & 0.10  & 0.00  & 0.07  & 0.24 \\\hline
{\bf Beet2} & 5.49 & 2.51 & 2.73 & 2.25 & 5.52 & 2.34 & 0.00 & 2.39 \\
 & 0.21  & 0.16  & 0.17  & 0.10  & 0.09  & 0.07  & 0.00  & 0.24 \\\hline\hline
{\bf Sch\"on} & 2.62 & 1.54 & 1.52 & 1.79 & 2.02 & 2.31 & 2.39 & 0.00 \\
 & 0.28  & 0.18  & 0.18  & 0.19  & 0.20  & 0.24  & 0.24  & 0.00 \\
\end{tabular}
\end{center}
\end{table}

\subsection{A thorough exposition of the tables}
These results yield many tables which do not fit this article and would make
this exposition clumsy.
Their thorough exposition are in~\cite{ksstats},
with all the tables and descriptions.

\subsection{Scripts for automated generation of the tables}
The tables that are benchmarks and that result from comparing real samples
are rendered by scripts.
These scripts are configurable, i.e. might be set to render other tables if needed.
Once the new tables are rendered, they might be assembled into a PDF by means
of the latex files that yield~\cite{ksstats}.

\section{CONCLUSIONS AND FURTHER WORK}\label{sec:conc}
This exposition described the $c'$ statistical distance,
the benchmark tables for $c'$ and its use to observe differences in real samples.
As far as I understand, the tables are effective in exposing reference
values in various settings of various distributions.
The Kolmogorv-Smirnov test, from which $c'$ is derived, is known
to be robust in the sense that it is usable even when the underlying distributions
are not known or present problems for other tests.
The overall result is that we obtained a statistical distance which is
useful in various contexts and have now benchmark tables.

Potential next steps are:
\begin{itemize}
	\item better organize the scripts that render the benchmark tables,
		because they are scattered in the \texttt{tests/} directory
		of~\cite{gmaneLegacy}.
	\item Make a better presentation of the benchmark tables in~\cite{ksstats}.
		They are sound but were made for personal usage and might be enhanced
		by better descriptions and contextualization.
	\item Use other distributions for obtaining the tables.
		This is relevant mainly because the number of rejections of the null hypothesis
		was sometimes higher that expected for the significance level in power-law distributions.
	\item Compare $c'$ to other statistical distances: in which cases are they suitable, preferable and
		what results they yield.
	\item Give a more formal account of the conditions needed for $c'$ to be considered a metric 
		and for the cases where $c'$ does not satisfy Equation (2).
	\item Obtain reference values of $c'$ for simulations where the null hypothesis is true.
\end{itemize}
\noindent Finally, the most urgent developments this contribution needs are:
1) a description of the differences in $c'$ in the cases of continuous and discrete distributions;
and 2) implement these measurements without using histograms because they are not needed to
attain the cumulative distribution used for the Kolmogorov-Smirnov statistic and because
$c'$ might be regarded as a metric (not a generalized metric) if obtained without using histograms,
as exposed in Section~\ref{sec:desc}.

\subsection*{\textit{Acknowledgements}}
The author thanks CNPq for the funding received while researching the topic of this article,
the researchers of IFSC/USP and ICMC/USP for the recurrent collaboration in every situation
where we needed directions for investigation.








\end{document}